# Vibrating wire alignment technique

WANG Xiao-Long(王小龙)[1], DONG lan(董岚)[1], WU lei(吴蕾)[1,2], LI Chun-Hua(李春华)[1]

[1]Institute of High Energy Physics, Chinese Academy of Sciences, Beijing 100049, China

[2]University of Chinese Academy of Sciences, Beijing 100049, China

**Abstract**: Vibrating wire alignment technique is a kind of method which through measuring the spatial distribution of magnetic field to do the alignment and it can achieve very high alignment accuracy. Vibrating wire alignment technique can be applied for magnet fiducialization and accelerator straight section components alignment, it is a necessary supplement for conventional alignment method. This article will systematically expound the international research achievements of vibrating wire alignment technique, including vibrating wire model analysis, system frequency calculation, wire sag calculation and the relation between wire amplitude and magnetic induction intensity. On the basis of model analysis this article will introduce the alignment method which based on magnetic field measurement and the alignment method which based on amplitude and phase measurement. Finally, some basic questions will be discussed and the solutions will be given.

## 1. Introduction

Vibrating wire aligntment technique was first put forward by Alexander Temnykh of Cornell University. Inspired by a kind of magnetic field measurement method 'pulsed-wire' [1], Alexander Temnykh put forward use vibrating wire to do the magnet alignment [2]. Compared with pulsed-wire, vibrating wire can substantially shorten the wire required for field measurement, which makes it become possible for applying vibrating wire for the long straight section components alignment in accelerator. Compared with pulsed-wire, another advantage of vibrating wire is that it does not need to generate a pulse of current to pass through the long wire, so it becomes more convenient for use.

Vibrating wire alignment technique is a kind of method which through measuring the spatial distribution of magnetic field to do the alignment, its theory is not same as the conventional alignmet method which based on the mechanical structure of magnet to do the alignment. Because it is through measuring the magnetic field to find the center and establish the relation between the center and the fiducials, vibrating wire method can substantially decrease the errors income and can achieve very high accuracy. According to Brookhaven National Laboratory (BNL) experiences vibrating wire method can align 6m straight section components within 0.03mm [3], this is impossible for conventional alignment method. Vibrating wire can be applied for magnet fiducialization, several magnets' center alignment which are arranged in line [4], magnet alignment in enclosed environment [5]. Vibrating wire method can achieve high accuracy in a relatively small range, if combined with conventional alignment method it will effectively improve alignment accuracy for the whole accelerator complex.

## 2. Vibrating wire model analysis

Vibrating wire system is shown in Fig. 1.

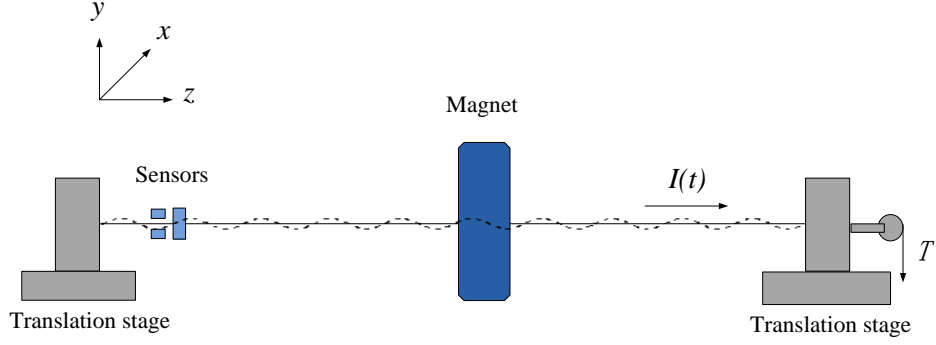

Fig. 1. Vibrating wire system.

A thin wire is through the center of the magnet, the ends of the wire are fixed on the motor-drived translation stages. The stages can move along *x* and *y* direction. The tension of the wire is T, a current *I(t)* is connected to the wire, at the end of the wire two sensors are used to measure the wire movement in *x* and *y* direction.

This system can be seen as a typical string forced vibration model, the driving force is a Lorenz force generated by the current and the magnetic field and the frequency of the driving force is determined by the frequency of the current. It can be analyzed according to the string forced vibration knowledge in physics [6].

The wire in this model connecting with alternating current will be affected by stress, gravity, damping force and Lorenz force. Firstly to analyze its movement in *y* direction. For each point in the wire, its *y* coordinate is a function of coordinate *z* and time *t*, write as *y(z,t)*. The analysis of stress for a section of wire of length d*z* is shown in Fig. 2.

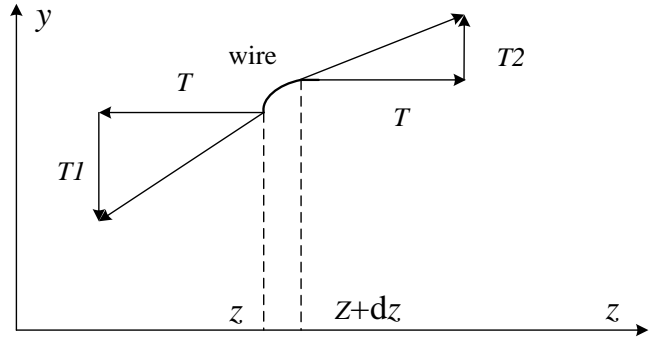

Fig. 2. The analysis of stress for a section of wire of length d*z*.

d*z* section will not move along *z* direction, so the total force in *z* direction is 0. In *y* direction, on the left side the force is $T_1 = T(\frac{\partial y}{\partial z})\big|_z$, on the right side the force is $T_2 = T(\frac{\partial y}{\partial z})\big|_{z+dz}$. Then for d*z* section the total force is $T_2 - T_1 = T(\frac{\partial y}{\partial z})\big|_{z+dz} - T(\frac{\partial y}{\partial z})\big|_z = T\int_z^{z+dz} \frac{\partial^2 y}{\partial z^2} dz$. When the length of d*z* is small enough it can be written as $T_2 - T_1 = T\frac{\partial^2 y}{\partial z^2} dz$. The gravity for d*z* section is $-\mu g dz$, where $\mu$ is the density of the wire and *g* is the acceleration of gravity coefficient. The damping of d*z* segment is $-\gamma \frac{\partial y}{\partial t} dz$, where $\gamma$ is the damping coefficient. The Lorenz force of d*z* segment is *I(t)B_x(z)*d*z*, where *I(t)* is the driving current, $B_x(z)$ is the magnetic induction intensity in $x$ direction. So the differential equation of motion for d$z$ segment is

$$\mu dz \frac{\partial^2 y}{\partial t^2} = T \frac{\partial^2 y}{\partial z^2} dz - \mu g dz - \gamma \frac{\partial y}{\partial t} dz + I(t) B_x(z) dz.$$

Dividing through by d$z$ and rearranging terms it will be

$$\mu \frac{\partial^2 y}{\partial t^2} + \gamma \frac{\partial y}{\partial t} - T \frac{\partial^2 y}{\partial z^2} = -\mu g + I(t) B_x(z). \quad (1)$$

This is a nonhomogeneous order two differential equations and will be discussed in the following.

### 2.1 Natural frequency of wire

when no forces act on the wire, from equation (1) can get

$$\mu \frac{\partial^2 y}{\partial t^2} + \gamma \frac{\partial y}{\partial t} - T \frac{\partial^2 y}{\partial z^2} = 0. \quad (2)$$

Because $z$ and $t$ are two independent variables, by separation of variables $y(z,t) = Y_z(z) Y_t(t)$, then get

$$\frac{1}{T Y_t}(\mu \frac{d^2 Y_t}{dt^2} + \gamma \frac{dY_t}{dt}) = \frac{1}{Y_Z} \frac{d^2 Y_Z}{dz^2}. \quad (3)$$

Since both sides are functions of different variables, they must be constants. Set this constant equal to $-k^2$. From the right side of equation (3) we can get

$$\frac{d^2 Y_Z}{dz^2} + k^2 Y_Z = 0. \quad (4)$$

Equation (4) is a two order homogeneous linear differential equation with constant coefficients, it'scharacteristic equationis

$$r^2 + pr + q = 0.$$

Where $p$=0, $q$=$k^2$. The solutions are $r=\pm ik$. So the solution of equation (4) is

$$Y_Z = C_1 \sin kz + C_2 \sin kz$$

with the boundary conditions $Y_z(0)=Y_z(L)=0$, we can get $C_1$=0, $k = n\pi/L$. So,

$$Y_Z(z) = C_2 \sin(\frac{n\pi z}{L}).$$

Where n=1,2,3⋯, $C_2$ is an arbitrary constant. From the left side of equation (3) we can get

$$\frac{d^2 Y_t}{dt^2} + \frac{\gamma}{\mu} \frac{dY_t}{dt} + \frac{T}{\mu}(\frac{n\pi}{L})^2 Y_t = 0. \quad (5)$$

Set $\alpha = \frac{\gamma}{\mu}$, $\omega_n^2 = \frac{T}{\mu}(\frac{n\pi}{L})^2$ we can get

$$\frac{d^2 Y_t}{dt^2} + \alpha \frac{dY_t}{dt} + \omega_n^2 Y_t = 0. (6)$$

This is a two order constant coefficientshomogeneous linear differential equation about $Y_t(t)$, its characteristic equation is

$$r^2 + \alpha r + \omega_n^2 = 0.$$

The solutions are a pair of conjugate complex roots, $r = -\frac{\alpha}{2} \pm i \sqrt{\omega_n^2 - (\frac{\alpha}{2})^2}$. So the solution of $Y_t$ is

$$Y_t = e^{-\frac{\alpha}{2}t}\left[c_1\cos\left(\sqrt{\omega_n^2 - \left(\frac{\alpha}{2}\right)^2}\,t\right) + c_2\sin\left(\sqrt{\omega_n^2 - \left(\frac{\alpha}{2}\right)^2}\,t\right)\right].$$

Where $c_1$, $c_2$ are constants. The general solution of equation (2) is

$$y(z,t) = Y_Z(z)Y_t(t) =$$

$$\sum_{n=1}^{\infty} e^{-\frac{\alpha}{2}t}\left[c_1\cos\left(\sqrt{\omega_n^2 - \left(\frac{\alpha}{2}\right)^2}\,t\right) + c_2\sin\left(\sqrt{\omega_n^2 - \left(\frac{\alpha}{2}\right)^2}\,t\right)\right]\sin(\frac{n\pi z}{L}). \quad (7)$$

From equation (7) it can be found when no forces act on the wire and when damping is zero ($\gamma = 0$), the natural frequency of the wire is $\omega_n$. Because the damping is very little, the natural frequency of the wire can be deemed as $\omega_n = 2\pi\frac{n}{2L}\sqrt{\frac{T}{\mu}}$ and the basic natural frequency is $\omega_1 = 2\pi\frac{1}{2L}\sqrt{\frac{T}{\mu}}$.

## 2.2 The sag of wire due to gravity

When only gravity act on the wire, from equation (1) we can get

$$\mu\frac{\partial^2 y}{\partial t^2} + \gamma\frac{\partial y}{\partial t} - T\frac{\partial^2 y}{\partial z^2} = -\mu g. \quad (8)$$

For gravity is an item has nothing to do with time, we can get

$$\frac{\partial^2 y}{\partial z^2} = \frac{\mu g}{T}.$$

After two times integration

$$y(z) = \frac{\mu g}{2T}z^2 + c_1 z + c_2.$$

With the boundary conditions $y(0) = 0$, $y(L) = 0$ we get

$$y(z) = \frac{\mu g}{2T}z(z - L). \quad (9)$$

Considering a more common condition that the wire is tilt, as shown in Fig. 3.

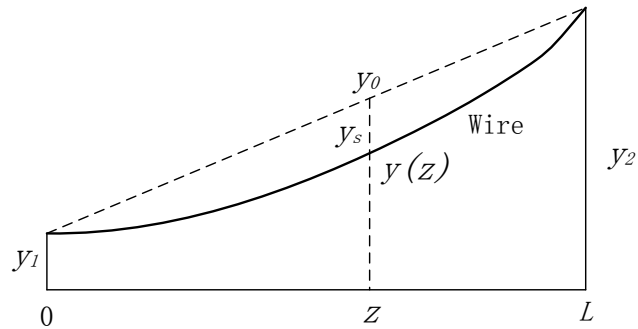

Fig. 3. Sage analysis of a tilt wire

The boundary condition becomes $y(0) = y_1$, $y(L) = y_2$ we can get

$$y(z) = \frac{\mu g}{2T}z(z - L) + \frac{y_2 - y_1}{L}z + y_1.$$

For

$$y_0(z) = \frac{y_2 - y_1}{L}z + y_1.$$

The sag

$$y_s = y_0(z) - y(z) = -\frac{\mu g}{2T}z(z - L).$$

This is as same as the level wire. Calculating the max sag S, according to equation (10), set $\frac{dy}{dz} = 0$ we can get at $z = \frac{L}{2}$ $S = \frac{\mu g L^2}{8T}$. For $\omega_1^2 = \frac{\pi^2 T}{\mu L^2}$ we can get $S = \frac{g\pi^2}{8\omega_1^2}$. For $\omega_1 = 2\pi f_1$, $f_1$ is the basic frequency, we can get $S = \frac{g}{32 f_1^2}$. Substitute that in equation (9) we can get

$$y(z) = \frac{g}{8 f_1^2 L^2} z(z - L). \quad (10)$$

Equation (10) is the sag equation, we can measure the $f_1$ by vibrating wire and calculate the sag to do sag correction.

### 2.3 The motion of wire driven by Lorenz force

When only Lorenz force act on the wire, from equation (1) we can get

$$\mu \frac{\partial^2 y}{\partial t^2} + \gamma \frac{\partial y}{\partial t} - T \frac{\partial^2 y}{\partial z^2} = I(t) B_x(z) \quad (11)$$

According to the phasor representation method of electric circuit theory, the current can be written as $I(t) = I_0 e^{i\omega t}$. According to physics knowledge, when the vibration get to steady state the vibrating frequency will become same as the driving force frequency [6], so we can set $y(z,t) = Y_B(z) e^{i\omega t}$ substitute that in equation (11) we can get

$$-\omega^2 \mu Y_B e^{i\omega t} + i\omega \gamma Y_B e^{i\omega t} - T \frac{d^2 Y_B}{dz^2} e^{i\omega t} = I_0 e^{i\omega t} B_x(z).$$

re-arrange it we can get

$$-\omega^2 \mu Y_B + i\omega \gamma Y_B - T \frac{d^2 Y_B}{dz^2} = I_0 B_x(z). \quad (12)$$

Equation (12) is a very complicated differential equation, $Y_B(z)$ and $B_x(z)$ are both functions of $z$ and the expression of them are very complicated. In order to easily solve equation (12), we can use the boundary condition $Y_B(0) = 0, Y_B(L) = 0$, expand $Y_B(z)$ and $B_x(z)$ in Fourier sine series.

$$Y_B(z) = \sum_{n=1}^{\infty} Y_{Bn} \sin\left(\frac{n\pi z}{L}\right). \quad (13)$$

where $Y_{Bn}$ is a constant.

$$B_x(z) = \sum_{n=1}^{\infty} B_{xn} \sin\left(\frac{n\pi z}{L}\right). \quad (14)$$

where $B_{xn}$ is a constant. Substitute them in equation (12) we can get

$$-\omega^2 \mu \left(\sum_{n=1}^{\infty} Y_{Bn} \sin\left(\frac{n\pi z}{L}\right)\right) + i\omega \gamma \left(\sum_{n=1}^{\infty} Y_{Bn} \sin\left(\frac{n\pi z}{L}\right)\right) + T \left(\sum_{n=1}^{\infty} Y_{Bn} \sin\left(\frac{n\pi z}{L}\right) \left(\frac{n\pi}{L}\right)^2\right) =$$

$$I_0 \left(\sum_{n=1}^{\infty} B_{xn} \sin\left(\frac{n\pi z}{L}\right)\right).$$

for arbitrary $n$

$$\left[-\omega^2 + i\omega \frac{\gamma}{\mu} + \frac{T}{\mu}\left(\frac{n\pi}{L}\right)^2\right] Y_{Bn} = \frac{1}{\mu} I_0 B_{xn}.$$

Set $\alpha = \frac{\gamma}{\mu}$, $\omega_n^2 = \frac{T}{\mu}\left(\frac{n\pi}{L}\right)^2$ we can get

$$Y_{Bn} = \frac{-I_0 B_{xn}}{\mu(\omega^2 - \omega_n^2 - i\omega\alpha)}.$$

Then

$$Y_B(z) = \sum_{n=1}^{\infty} \frac{-I_0 B_{xn}}{\mu(\omega^2 - \omega_n^2 - i\omega\alpha)} \sin\left(\frac{n\pi z}{L}\right).$$

According to the complex solution method of vibration differential equation in physics, $y(z,t)$ is equal to the real part of $Y_B(z)e^{i\omega t}$.

$$y(z,t) = \mathrm{Re} \sum_{n=1}^{\infty} \frac{-I_0 B_{xn}}{\mu(\omega^2 - \omega_n^2 - i\omega\alpha)} \sin\left(\frac{n\pi z}{L}\right) e^{i\omega t} =$$

$$\sum_{n=1}^{\infty} \frac{-I_0 B_{xn} \sin\left(\frac{n\pi z}{L}\right)}{\mu[(\omega^2 - \omega_n^2)^2 + (\omega\alpha)^2]} [(\omega^2 - \omega_n^2)\cos(\omega t) - \omega\alpha\sin(\omega t)]. \quad (15)$$

The motion of wire in $x$ direction is similar to that in $y$ direction, the differences in $x$ direction are no gravity act on the wire and the Lorenz force is generated by $B_y(z)$. So it can be seen the natural frequency in $x$ direction is same as that in $y$ direction and don't need to care about the gravity. The equation of Lorenz force in $x$ direction can be get by substitute $B_{xn}$ to $B_{yn}$ in equation (15). Until here we finish the vibrating wire model analysis.

3. **Magnet alignment**

Equation (15) describe the relation between the motion of point $z$ in $y$ direction and the magnetic induction intensity $B_{xn}$, but the description of equation (15) still is too complex and difficult to be applied for actual measurement. It needs to be further simplified for practical application.

3.1 **Magnet alignment based on the distribution of magnetic field measurement**

Alexander Temnykh provided a method which to do magnet alignment by measuring the distribution of the magnetic induction intensity in $x$-$y$ cross section [2,7]. Take an example for $y$ direction, we can construct a function

$$\mathcal{F}(\omega) = \frac{1}{T}\int_0^T y(z,t)\,\mathrm{Re}(I_0 e^{i\omega t})\mathrm{d}t \quad . (16)$$

Where $T$ is the sampling time, it should be integral multiple of the current cycle. $y(z,t)$ is the amplitude of point $z$ on the wire in $y$ direction. $I_0 e^{i\omega t}$ is the current. $\mathcal{F}(\omega)$ can be achieved by measurement. Set the $z$ to be the position of the sensor, through the sensor sampling the motion data of wire in y direction during $T$ period and sampling the current at the same time, use these to do numerical integration as shown in equation (16). The sampling rate can be selected according to the fastest sampling rate of the instrument.

In order to measure the distribution of the magnetic induction intensity in $x$-$y$ cross section, the wire should be used to do scan measurement in $x$ and $y$ direction respectively. Take an example for $y$ direction, firstly we should use conventional alignment method align the wire to the center of the magnet, the accuracy can be 0.1mm, we can sure the center of magnet must be within 1mm from the wire. Secondly, take the wire current position as center and select a series of points in a 2mm wide range in y direction, at each point to do frequency scan measurement. When doing frequency scan measurement, we should set one natural frequency as center and use a series of currents which frequencies are close to it to measure the $\mathcal{F}(\omega)$. Then, we can get a graph of $\mathcal{F}(\omega) - \omega$. Fig. 4 is the scan result at one point.

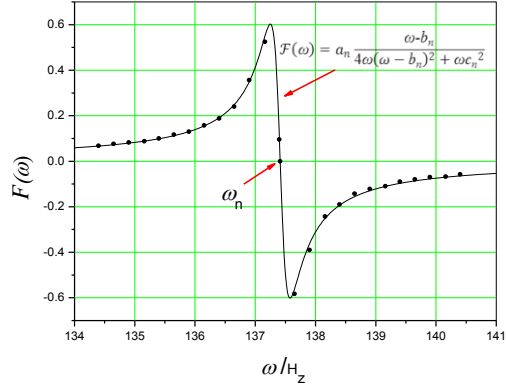

Fig. 4. Frequency scan

Through frequency scan we can get the magnetic induction intensity of this point. Substitute the $y(z,t)$ in equation (16) with equation (15) we can get

$$\mathcal{F}(\omega) = \sum_{n=1}^{\infty} \frac{-I_0{}^2 B_{xn} \sin(\frac{n\pi z}{L})(\omega^2-\omega_n^2)}{2\mu[(\omega^2-\omega_n^2)^2+(\omega\alpha)^2]} = \sum_{n=1}^{\infty} \frac{-I_0{}^2 B_{xn} \sin(\frac{n\pi z}{L})(\omega+\omega_n)(\omega-\omega_n)}{2\mu[(\omega+\omega_n)^2(\omega-\omega_n)^2+(\omega\alpha)^2]} \ .$$

When $\omega$ is near to $\omega_n$ it can be simplified as

$$\mathcal{F}(\omega) = \sum_{n=1}^{\infty} \frac{-I_0{}^2 B_{xn} \sin(\frac{n\pi z}{L}) 2\omega(\omega-\omega_n)}{2\mu[4\omega^2(\omega-\omega_n)^2+(\omega\alpha)^2]} = \sum_{n=1}^{\infty} \frac{-I_0{}^2 B_{xn} \sin(\frac{n\pi z}{L})(\omega-\omega_n)}{\mu[4\omega(\omega-\omega_n)^2+\omega\alpha^2]} \ . (17)$$

Set $\mathcal{F}(\omega) = \sum_{n=1}^{\infty} \mathcal{F}_n(\omega)$, Where

$$\mathcal{F}_n(\omega) = \frac{-I_0{}^2 B_{xn}}{\mu} \sin\left(\frac{n\pi z}{L}\right) \frac{\omega-\omega_n}{4\omega(\omega-\omega_n)^2+\omega\alpha^2} \ .(18)$$

For vibrating wire system is a weak damping system, $\frac{\omega\alpha^2}{\omega-\omega_n}$ will become very small in a certain range, for equation (18) when $\omega \approx \omega_n$, $\mathcal{F}_n(\omega)$ will become big notably, but when $\omega=\omega_n$, $\mathcal{F}_n(\omega) = 0$. This can be seen from figure 4, along with $\omega$ near to $\omega_n$, $\mathcal{F}(\omega)$ becomes big notably, that means $\mathcal{F}_n(\omega)$ is much bigger than other items, so when $\omega \approx \omega_n$, $\mathcal{F}(\omega) \approx \mathcal{F}_n(\omega)$.

We can simplify equation (18) as follow

$$\mathcal{F}(\omega) = a_n \frac{\omega-b_n}{4\omega(\omega-b_n)^2+\omega c_n^2} \ .(19)$$

Where $a_n$, $b_n$, $c_n$ are coefficients need to be solved. Compare with (18) can get $a_n = \frac{-I_0{}^2 B_{xn}}{\mu} \sin\left(\frac{n\pi z}{L}\right)$, $b_n = \omega_n$, $c_n = \alpha$. So after solve $a_n$ we can calculate $B_{xn}$. Using (19) as the model equation and use the measured data to do nonlinear fitting we can get $a_n$. Before doing the fit we need to set the initial values for $a_n$, $b_n$, $c_n$, the $b_n$ initial value can be calculated by equation $\omega_n = 2\pi \frac{n}{2L}\sqrt{\frac{T}{\mu}}$. In order to calculate the initial values for $a_n$ and $c_n$ we need to rearrange (19)

$$\frac{1}{\mathcal{F}(\omega)} = x_1 \times 4\omega^2 - x_2 \times 4\omega + x_3 \times \frac{\omega}{\omega-b_n} \ . (20)$$

Where $x_1 = \frac{1}{a_n}$ $x_2 = \frac{b_n}{a_n}$ $x_3 = \frac{c_n^2}{a_n}$. Substitute $b_n$ with initial value in (20) through linear fitting we can solve $x_1$ $x_2$ $x_3$ then calculate the initial values for $a_n$ and $c_n$. The result of nonlinear fitting is shown in Fig. 4. According to (14), if we measure many orders of $B_n$ we can approximately calculate the $B_x$ at the wire location.

After doing frequency scan at every point we will get the distribution of magnetic induction intensity in *y* direction as shown in Fig. 5. From Fig. 5 it can be found the magnetic center is at 0.1mm. The measurement of magnetic induction intensity in *x* direction is same as *y*, the differences are to use the sensor measure $x(z,t)$ and the magnetic induction intensity is $B_y(z)$.

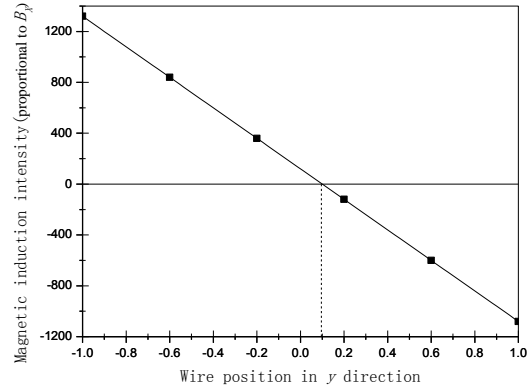

Fig. 5. Distribution of magnetic induction intensity

For magnet alignment it needs not only to align the magnetic center but also to align the roll pitch and yaw. Vibrating wire cannot measure roll, so we should use conventional method to do roll alignment, but vibration wire can measure pitch and yaw. According to (14), through measuring many orders $B_n$ we can approximately calculate the $B(z)$ and the relation between $B(z)$ and quadrupole position is shown in figure 6.

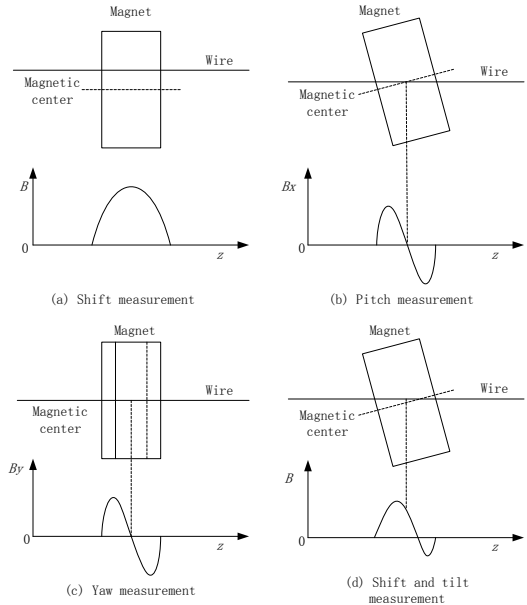

Fig. 6. Relation between $B(z)$ and magnet position

When pitch and yaw are 0 and magnetic center is on the wire, the $B(z)$ should be 0. When pitch and yaw are 0 and magnetic center is not on the wire, the $B(z)$ should be as shown in Fig. 6.(a), there is a peak at the magnet center location. When only pitch is not 0, the $B_x(z)$ should be as shown in Fig. 6.(b), there are two symmetrical peaks of different polarities at the magnet location. When only yaw is not 0, the $B_y(z)$ should be as shown in Fig. 6.(c), there are two symmetrical peaks of different polarities at the magnet location. When title is not 0 and shift is not 0, the $B(z)$ should be as shown in Fig. 6.(d), there are two unsymmetrical peaks of different polarities at the magnet location. One thing need to be noticed is the accuracy of $B(z)$ is determined by the highest order $B_n$, the higher of the order the shorter of the wavelength and the more accuracy of the description of the $B(z)$. In Fig.6 if the wavelength of the highest order $B_n$ is longer than the magnet it will not get the result in Fig.6.

### 3.2 Magnet alignment based on the amplitude and phase measurement

ZacharyWolf provided a method which to do quadruple alignment by measuring wire amplitude and phase [8]. According to the forced vibration knowledge in physics, for week damping system, when the drive force frequency $\omega$ is same to one of the system natural frequency $\omega_k$ the system will generate resonance [6]. When resonance the $n=k$ item in equation (15) will be much bigger than other items and for any point of the wire it's motion phase will be lag of the drive force $\frac{\pi}{2}$, so (15) can be simplified as

$$y(z,t) \approx \frac{I_0 B_{xn}}{\mu \omega_n \alpha} \sin\left(\frac{n\pi z}{L}\right) \cos\left(\omega_n t - \frac{\pi}{2}\right). \quad (21)$$

Take an example for $y$ direction, when we begin to find the magnetic center of magnet we first use conventional alignment method align the wire to the center of magnet and select a natural frequency $\omega_n$ as the driving current frequency. In order to get the actual resonance frequency, it need to monitor the wire motion phase and the driving current phase, adjust the current frequency until the wire phase is $\frac{\pi}{2}$ lag of the current phase, use a lock-in amplifier to get it. For quadruple there is $B_x = G y_d$, where $y_d$ is the distance between wire and magnetic center in $y$ direction. To get the value of $y_d$, we need to deduce the relation between $y_d$ and $B_{xn}$ and substitute it into (22) to solve.

According to Fourier transformation there is

$$B_{xn} = \frac{2}{L} \int B_x(z) \sin\left(\frac{n\pi z}{L}\right) dz .$$

Set the effective length of magnetic field $L_m$, the magnetic center location in $z$ direction $z_m = \frac{L}{j}$. When the tilt of magnet relative to wire is very small, the $B_x(z)$ can be seen as a constant $B_x$, then

$$B_{xn} = \frac{2}{L} \int_{\frac{L}{j}-\frac{L_m}{2}}^{\frac{L}{j}+\frac{L_m}{2}} B_x \sin\left(\frac{n\pi z}{L}\right) dz = -\frac{2}{L} B_x \frac{L}{n\pi} \cos\left(\frac{n\pi z}{L}\right)\bigg|_{\frac{L}{j}-\frac{L_m}{2}}^{\frac{L}{j}+\frac{L_m}{2}} = \frac{4B_x}{n\pi} \sin\left(\frac{n\pi}{j}\right) \sin\left(\frac{n\pi}{2}\frac{L_m}{L}\right) =$$

$$\frac{4G y_d}{n\pi} \sin\left(\frac{n\pi}{j}\right) \sin\left(\frac{n\pi}{2}\frac{L_m}{L}\right). \quad (22)$$

Substitute into (21)

$$y(z,t) \approx \frac{I_0 4 G y_d}{\mu \omega_n \alpha n\pi} \sin\left(\frac{n\pi}{j}\right) \sin\left(\frac{n\pi}{2}\frac{L_m}{L}\right) \sin\left(\frac{n\pi z}{L}\right) \cos\left(\omega_n t - \frac{\pi}{2}\right).$$

So the relation between the maximum amplitude $y_{max}(z)$ and $y_d$ is

$$y_{max}(z) = \left[\frac{I_0 4G}{\mu \omega_n \alpha n\pi} \sin\left(\frac{n\pi}{j}\right) \sin\left(\frac{n\pi}{2}\frac{L_m}{L}\right) \sin\left(\frac{n\pi z}{L}\right)\right] |y_d| . \quad (23)$$

Using (23) the magnetic center in *y* direction can be calculated by measuring the $y_{max}(z)$.

*X* direction magnetic center measurement is same as *y* direction, the differences are use sensor measures the *x(z,t)* and the magnetic induction intensity is $B_y(z)$.

For quadruple, the directions of magnetic induction intensity on both sides of the magnet center are opposite as shown in Fig. 7.

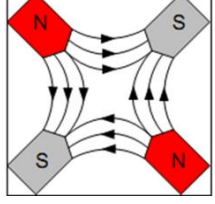

Fig. 7. Quadruple magnetic field

Correspondingly, the phase of wire relative to the driving current will also change, from $-\frac{\pi}{2}$ to $\frac{\pi}{2}$ or from $\frac{\pi}{2}$ to $-\frac{\pi}{2}$. By monitor the wire phase change we can judge on which side the wire is. Take an example for *y* direction, the relation of the wire motion phase $\emptyset$ and $y_d$ is

$$\emptyset = -\text{sign}(y_d)\left(\frac{\pi}{2}\right)$$

When the tilt of magnet relative to the wire is not small and need to do tilt alignment, the $B(z)$ at wire location will not be seen as a constant, it should be expressed as a function of the tilt $\theta$. Take an example for quadruple pitch, after finish the alignment in *x*、*y* direction to do the pitch alignment as shown in Fig.8

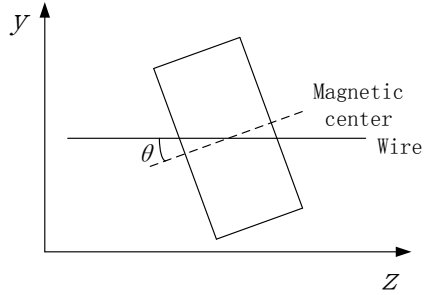

Fig. 8. Quadruple pitch alignment

$$B_x(z) = -G\left(z - \frac{L}{j}\right)\tan\theta \approx -G\left(z - \frac{L}{j}\right)\theta$$

$$B_{xn} = \frac{2}{L}\int_{\frac{L}{j}-\frac{L_m}{2}}^{\frac{L}{j}+\frac{L_m}{2}} B_x(z)\sin\left(\frac{n\pi z}{L}\right)dz = \frac{2}{L}\int_{\frac{L}{j}-\frac{L_m}{2}}^{\frac{L}{j}+\frac{L_m}{2}} -G\left(z - \frac{L}{j}\right)\theta\sin\left(\frac{n\pi z}{L}\right)dz$$

$$= \frac{2G}{L}\left[\frac{L_m L}{n\pi}\cos\frac{n\pi}{j}\cos\frac{n\pi L_m}{2L} - 2\left(\frac{L}{n\pi}\right)^2\cos\frac{n\pi}{j}\sin\frac{n\pi L_m}{2L}\right]\theta$$

Substitute into (21)

$$y(z,t) \approx \frac{I_0 2G}{\mu\omega_n\alpha L}\left[\frac{L_m L}{n\pi}\cos\frac{n\pi}{j}\cos\frac{n\pi L_m}{2L} - 2\left(\frac{L}{n\pi}\right)^2\cos\frac{n\pi}{j}\sin\frac{n\pi L_m}{2L}\right]\theta\sin\left(\frac{n\pi z}{L}\right)\cos\left(\omega_n t - \frac{\pi}{2}\right)$$

So the relation of $y_{max}(z)$ and $\theta$ is

$$y_{max}(z) = \frac{I_0 2G}{\mu \omega_n \alpha L} \left[ \frac{L_m L}{n\pi} \cos\frac{n\pi}{j} \cos\frac{n\pi L_m}{2L} - 2\left(\frac{L}{n\pi}\right)^2 \cos\frac{n\pi}{j} \sin\frac{n\pi L_m}{2L} \right] \sin\left(\frac{n\pi z}{L}\right) |\theta| \quad .(24)$$

The relation between wire motion phase relative to the driving current $\emptyset$ and $\theta$ is

$$\emptyset = \text{sign}(\theta)\left(\frac{\pi}{2}\right)$$

For $\theta > 0$, $\emptyset = \frac{\pi}{2}$ and $\theta < 0$, $\emptyset = -\frac{\pi}{2}$. Yaw measurement method is same as pitch, the differences are use sensor measures the $x(z,t)$ and the magnetic induction intensity is $B_y(z)$.

4. **Some basic things**

    **4.1 Sensor**

    The sensor used in vibrating wire is a kind of optical chopper as shown in Fig.9.

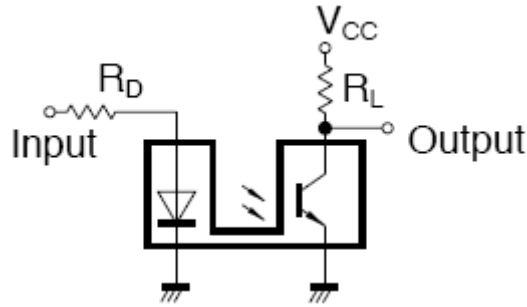

Fig. 9. Optical chopper

It's output character is shown in Fig.10.

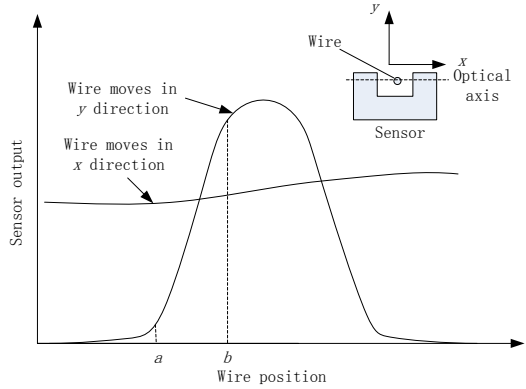

Fig. 10. Sensor output character

When the wire moves between a and b it's output and motion have a linear relation. So we should control the amplitude of wire at the sensor position, make it smaller than the distance between a and b. Sensor output is sensitive to the wire move direction, so we need use two sensors to monitor the wire motion in $x$、$y$ respectively.

**4.2 Wire**

The wire used in vibrating wire system must be no magnetic, we can choose Cu-2% Be wire. The diameter of wire can be about 0.1mm, too big will increase the damping, too small will decrease its tensile strength.

**4.3 Driving current frequency**

According to the content of 3.1, by measuring the magnetic induction intensity components we can calculate the magnetic induction intensity, but we do not need to measure many components, just measure some most representative components. From equation (14) it can be seen when $n = \frac{(1+2k)L}{2z_m}$ （$k$=0,1,2…，$z_m$ is magnet $z$ coordinate）the $n$ order component is the biggest one at the magnet location, so we can according this to choose the driving current frequency.

According to equation (22) when $z_m = \frac{(1+2k)L}{2n}$ （$k$=0,1,2…，$z_m$ is magnet $z$ coordinate）$B_{xn}$ gets to its biggest value, from equation (21) the wire's biggest amplitude $y_{max}(z)$ is proportional to $B_{xn}$, so according to the location of magnet to choose the driving current frequency can improve the measurement accuracy. In conclusion, we should make the magnet located at the peak of the wire $n$ order vibrating mode.

### 4.4 Sag correction

The nominal reference of vibrating wire is the line between the ends of the wire, but in actual the wire is a catenary because of the gravity, we need to do sag correction. According to (10) by measuring the basic frequency we can calculate the sag at the magnet center, after subtract the sag value from the magnetic center value in $y$ direction we can get the distance between the magnetic center and the nominal reference.

### 4.5 Background fields elimination

Vibrating wire alignment technique is based on magnetic field measurement, but in actual measurement the magnetic fields act on the wire are not only come from the magnet under measurement, they can come from other magnets at near and the earth field. In order to eliminate these magnetic fields we can first use vibrating wire to do a scan measurement within the measurement range when the magnets are off power. By doing this we can get the background fields distribution. Through best-fit we can get the magnetic induction intensity distribution curves in $x$ and $y$ direction relative to the wire position. Then, make the magnet need to be aligned on power and do the scan measurement again, through best-fit we can get the distribution curves under the magnet on power. The intersection of the on power curve and the off power curve is the magnetic center that has been eliminated the background fields.

## 5. Conclusions

Vibrating wire alignment technique is a kind of high accuracy alignment technique, it can be applied for small range straight section components alignment or components fiducialization, it is a necessary supplement for conventional alignment method. In this article we analyzed the vibrating wire mode in detail, deduced the resonance frequency equation, sag equation and wire amplitude and magnetic induction intensity relation equation, after that gave two kinds of alignment method, at last discussed some basic things and gave the solutions.